**Is Third-Party Provided Travel Time Helpful to Estimate Freeway Performance Measures?**

**Sakib Mahmud Khan**
California Partners for Advanced Transportation Technology
Institute of Transportation Studies
University of California, Berkeley
Berkeley, CA 94720
Email: sakibk@berkeley.edu

**Anthony David Patire**
California Partners for Advanced Transportation Technology
Institute of Transportation Studies
University of California, Berkeley
Berkeley, CA 94720
Email: adpat@berkeley.edu

Word Count: 5,510 words + 3 table (250 words per table) = 6,260 words

*Submitted 7/31/2020*

## ABSTRACT

Transportation agencies monitor freeway performance using various measures such as VMT (Vehicle Miles Traveled), VHD (Vehicle Hours of Delay), and VHT (Vehicle Hours Traveled). Public transportation agencies typically rely on point detector data to estimate these freeway performance measures. Point detectors, such as inductive loops cannot capture the travel time for a corridor, which can lead to inaccurate performance measure estimation. This research develops a hybrid method, which estimates of freeway performance measures using a mix of probe data from third-parties and data from traditional point detectors. Using a simulated I-210 model, the overall framework using multiple data sources is evaluated, and compared with the traditional point detector-based estimation method. In the traditional method, point speeds are estimated with the flow and occupancy values using the g-factors. Data from 5% of the total vehicle are used to generate the third-party vendor provided travel time data. The analysis is conducted for multiple scenarios, including peak and off-peak periods. Findings suggest that fusing data from both third-party vendors and point detectors can help estimate performance measures better, compared to the traditional method, in scenarios that have noticeable traffic demand on freeways.

**Keywords:** Third-party, travel time, performance measure, VMT, VHT, VHD

**INTRODUCTION**

Transportation agencies need to monitor freeway performance. For example, in California, every district within the department of transportation generates a quarterly report, called a Mobility Performance Report or MPR, where congestion information is reported (*1*). To compute VHD or Vehicle Hours of Delay and VMT or Vehicle Miles Traveled, data from VDS (Vehicle Detector Stations) are used which are available from California Performance Measurement System (PeMS). A VDS is a group of detectors located at a certain position on the freeway in a particular direction. In California, the overall VDS network includes 40,000 individual detection zones (*2*), and maintaining such a vast infrastructure requires extensive operational and maintenance support. Some data to calculate the performance measures can be obtained from third-party vendors. Typical data includes a report date, timestamp, link identifier, link length, speed, and travel time (*3*). Data from these vendors can augment data captured by public agencies to compute performance measures (*3*, *4*). There exists an opportunity to improve the accuracy of performance measures by fusing data from both third-party vendors and traditional point detectors. The inherent characteristics of the two different data sources make the hybrid calculation a challenging task. The third-party provided data have to be projected on the same coordinate system to be fused with the VDS data. However, existing disparities between coordinate systems used by the vendors and transportation agencies can hinder the desired projection. These differences can arise from a change in linear reference systems, freeway segmentation definitions, data coverage, and roadway geometries (*3*). Third-party data does not typically disclose the available probe vehicle penetration levels. For freeways with mainline and HOV (High Occupancy Vehicle) lanes, only aggregated travel time information is provided, which does not distinguish travel times between the mainline and HOV lanes. However, point detectors like VDS provide lane-specific data, and if single loop detectors are used only flow and occupancy values are available for the related lane.

The objective of this research is to define a framework for freeway performance measure computation that can employ a mix of data from multiple data sources. The overall framework of using multiple data sources will be referred to as the hybrid method in the rest of this paper. While doing so, the following tasks are included:

1. Determine an efficient algorithm for including third-party data in the calculation of performance measures
2. Evaluate the effect of data mix on VMT, VHT, and VHD estimation:
    a. using third-party data
    b. using traditional detector data from VDS
    c. using erroneous and, limited data. For example, the traditional method calculates speeds from VDS using a g-factor method, which can result in erroneous speed estimates. Another example is the limited penetration level of probe vehicles from which the third-party generated travel times and speeds.

The hybrid method incorporates three steps. **Step 1:** the data is acquired from both traditional point detectors or VDS, and third-party vendors. An initial data quality check is conducted to evaluate whether or not the data is usable to estimate performance measures. **Step 2:** both flow and travel time data are conflated to project them onto the desired cell. Here, conflation means the projection of data from certain points on one map to other desired points (corresponding points) on a different map. The term cell is used to denote the desired domain of analysis in the hybrid method, as shown in **Figure 1**. A cell is a small length of freeway used to perform a fine level of analysis to narrow down areas of congestion. **Step 3:** After having both data conflated, data fusion is performed to calculate the desired performance measures. The term fusion indicates the final integration of flow (from a traditional source) and travel time data (from a third-party).

**Figure 1** presents the data conflation and fusion steps with a schematic diagram. It shows that data from VDS (**Figure 1a**) are available on specific points along the freeway, which does not, in general, line up with the layout of the third-party vendor-provided data (**Figure 1b**). Both VDS and third-party data can include imputed data, in case the real-time measured data is unavailable due to detector malfunction, communication error, or absence of probe vehicles. Often the imputed data are drawn from historic observations over the same spatio-temporal domain. Once data are conflated on the same network (**Figure 1c**), both flow and travel time data are available for each cell. Later using the conflated data, performance measures are estimated on the cells and aggregated over the total spatio-temporal coverage of interest.

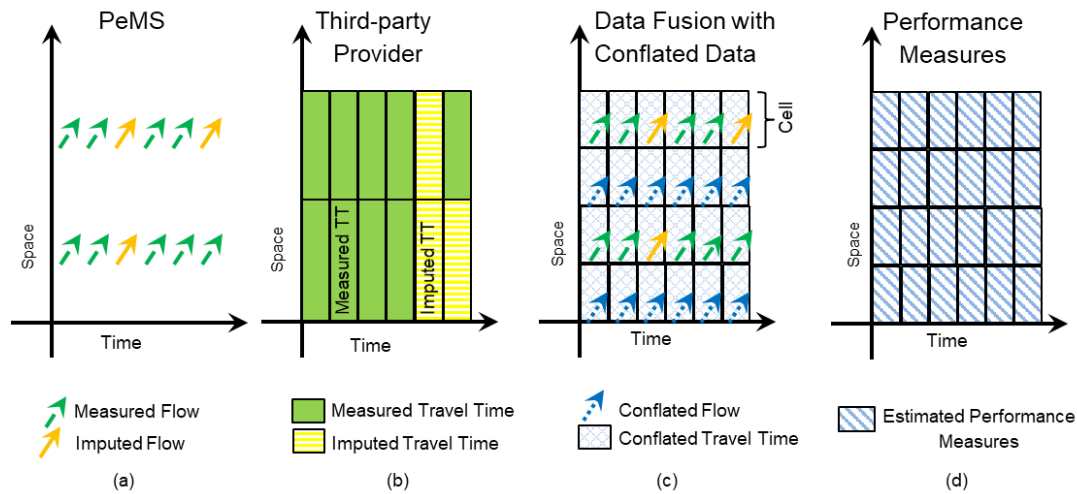

**Figure 1 Data conflation and fusion**

The purpose of the research is to understand the accuracy improvements that are possible when incorporating third-party data into performance measures for freeway mainlines. Public transportation agencies can use this hybrid method to estimate freeway performance measures for any freeway mainline sections in their jurisdiction. This is intended to be useful to these agencies when deciding whether the improvement is worthwhile for the cost of the data. In this research, the evaluations of the developed algorithms are performed using a simulation model. Using any simulated network, the overall framework can be implemented. The simulation model generates data based on the data characteristics of point-detectors and any representative third-party vendor. In this research, a single vendor is considered. The following sections discuss the related studies, research method, research considerations, analysis and findings, and conclusions based on the research.

**RELATED STUDIES**

Data conflation and fusion have been studied in earlier literature (*5–14*). One practical problem for using third-party data from multiple vendors is that in general, data are collected, processed, and delivered on incompatible spatial reference systems. Maps from multiple parties tend not to align perfectly. Therefore, this review focuses on studies to interpolate or reconstruct a more complete picture of the traffic state from partial data. Also studies that fused data from multiple sources are reviewed. However, none of the studied research investigated the fusion of point detector data with the third-party vendor provided travel time data.

One study (*7*) presented a smoothing method that can interpolate data from single stationary sensors to intermediate points on a spatio-temporal domain. Data were reconstructed from incomplete information to identify bottlenecks efficiently. A non-linear weight-based reconstruction method used both congested and free-flow velocity information as the a priori traffic estimate. Two linear anisotropic kernel functions were used to smooth the available data based on traffic speed propagation in free-flow and congested regimes. Fixed values were used for a spatio-temporal smoothing window, and a perturbation propagation velocity and transition velocity. The method was used to identify bottlenecks in two real-world scenarios. Speeds were reconstructed for a freeway section using only 35% of the information. However, only visual comparison was used to evaluate the reconstructed data.

Another study (*6*) extended the method in (*7*) to provide a heuristic data fusion model that follows traffic flow theory and can reconstruct the data with inherent structural ambiguity in the spatio-temporal domain. The Extended Generalized Treiber Helbing filter was developed to align data for use with first-order traffic flow models and Kalman filters. An area-based restriction was applied to reconstruct data at any specific point. To fuse data from multiple data sources, a linear formulation was developed with one additional weight. This weight expressed the reliability of the data sources. A simulation study was used with a 19-km freeway segment in Europe with on-ramp, off-ramp and weaving sections. Compared to the single data source based reconstruction, bias was reduced while using data from multiple sources. Reconstruction error occurred at the edges of the congested region with wider detector spacing, lower floating car penetration, and coarser data from AVI. According to the authors, future extensions of this research could include better estimation of the model parameters from automated processes, or a priori estimates based on Bayesian statistics.

GASM (Generalized Adaptive Smoothing Method) was developed in (*5*) to incorporate data from heterogeneous sources. The motivations were to address the sparseness and noise of a single data source by combining data from multiple sources. A 12-km highway with 4 loop detectors and 10 floating cars was simulated. Using point speeds from a few floating cars along with the detector-based speeds, the smoothed velocity reconstruction achieved better accuracy compared to single-source based reconstructions. The method is applicable for freeways with detector spacings up to 1.8 miles. The close positioning of detectors was recommended for bottleneck locations to accurately identify areas of congestion. The generalized smoothing method is used by later studies (*8*).

One extension of the study in (*6*) was performed in (*9*) where a fusion algorithm was developed for urban expressways. The weight function was modified to fuse data from multiple sources. Real-world data from a 10-km corridor in Beijing was used. The data collection interval for the loop detectors was 2 min, while for GPS-based vehicles it was 5 min. Using only vehicle data, a minimum 5% penetration rate was required to provide a reliable estimate of travel times. In addition, the fusion of loop detector data and GPS data (at 2% penetration) outperformed travel time estimation from a single data source.

In (*13*), travel times were estimated using point-based GPS data from probe and loop detector data. One challenge was to map probe-based GPS points to the freeway. The authors used a Path Inference Filter for the projection. Selections of probe data were filtered and fed into a data fusion engine. The fusion methods used the Godunov scheme discretized Lighthill and Whitham model to estimate speed. Data were fused using EnKF (Ensemble Kalman Filtering). An iterative calibration process was used to fine-tune the model parameters. The EnKF was calibrated with Bluetooth based data. The model was validated with multiple networks. One key result was the relationship between GPS sample-rate and vehicle penetration rate. On freeways, better travel time estimation was achieved using data with a low sample-rate and a high penetration rate, than that achieved using data with a high sample-rate and a low penetration rate. In addition, where no loop detectors existed, travel time could be estimated with reliable accuracy using only probe vehicle data.

Data from multiple sources were used to reconstruct traffic in (*15*). A framework was developed to separate noise from data and to measure relationships among multiple data sources using the fundamental diagram. The Alternating Direction of Multipliers optimization method was used to obtain the reconstructed data. Real-world average speeds from 28 links using cell phones and floating cars were used for single parameter reconstruction. For multi-parameter reconstruction, real-world speeds, occupancies, and volume data from microwave detectors were used. For single-parameter reconstruction, with 80% data loss, the reconstruction error was less than 15%. However, this method is only applicable to uninterrupted flow.

In (*16*), freeway density was estimated from loop and probe vehicle data using a Rao-Blackwellized particle filter and stochastic cell transmission model to match the solution of partial differential equations with the available sensor data. One main assumption was that traffic density and velocity at any particular moment is independent of what it was at the previous moment.

## METHOD

This section discusses the research method used for the estimation of VMT, VHT, and VHD using both the traditional point-detector method and the hybrid method. In the end, this section discusses the evaluation setup and scenarios to evaluate the frameworks.

### Traditional Method

This sub-section describes the traditional method to calculate performance measures using only data from dedicated point-detectors such as loops in the freeway.

#### *Traditional Speed Estimation*

Using single loop detectors, flow (i.e., the number of vehicles passing a detector during a certain time interval) and occupancy (i.e., the percentage of the time during which the detector is occupied) values are captured. To calculate speed from flow and occupancy, another parameter called a g-factor is estimated which is the average effective length of vehicles passing over the detector. Assume that for any time interval $i$, $o_i$ and $q_i$ are the occupancy and flow values, respectively, for the loop detector. Using multiple iterations with the experimental setup, suitable g-factor values for each lane on the freeways are estimated. The preliminary speed, $s_i$, is:

$$s_i = \frac{g \cdot q_i}{o_i} \tag{1}$$

Here $g$ is the g-factor value. Using an exponential filter, the final calculated speed estimate $v_i$ is obtained from the estimated speed $s_i$, as shown in **Equation (2**. The variable, $w_i$, can be estimated with **Equation (3**. The value of the smoothing parameter, $a$, is considered from (*17*).

$$v_i = w_i \cdot s_i + (1 - w_i) \cdot v_{i-1} \tag{2}$$

$$w_i = \frac{q_i}{q_i + a} \tag{3}$$

The final calculated speed represents a point-speed for the associated loop detector. At any particular location, several individual loop detectors can form a VDS as shown in **Figure 2**. To estimate the VDS speed at time interval $i$, speed estimates from all loop detectors are averaged.

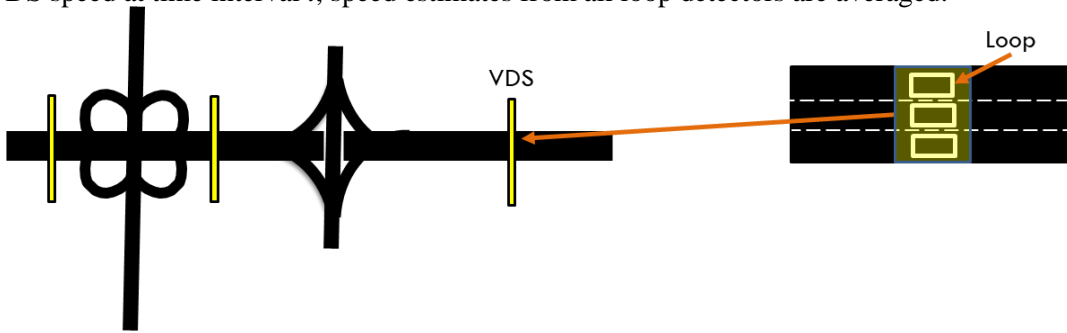

**Figure 2 Schematic of multiple loops forming single VDS**

#### *Traditional Performance Measures Estimation*

A typical assumption is that each VDS is representative of the freeway segment from the upstream midpoint to the downstream midpoint of neighboring VDS. In **Figure 3**, each green double arrow denotes the freeway segment of the VDS that lies within the segment. The yellow stars mark the midpoint of two successive VDS. In terms of traffic data, the key measurements captured or estimated at the VDS locations are counts, occupancies, and speeds.

For a particular time interval *i*, VMT is the sum of the total miles driven by all vehicles for a freeway in that time interval. VMT is represented in the units of vehicle-miles and can be calculated over a specific interval by the following equation.

$$\text{VMT} = \sum_i L \cdot q_i \tag{4}$$

where $q_i$ is the number of vehicles that passed over the VDS and *L* is the length of its associated freeway segment.

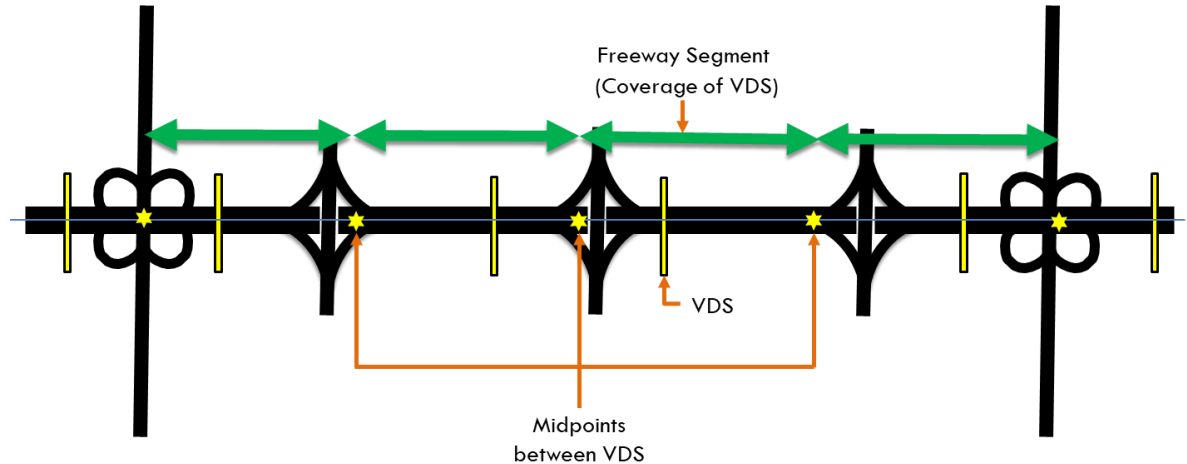

**Figure 3 Freeway segment representing coverage of VDS**

For any particular time interval *i*, VHT is the sum of the total hours driven by all vehicles for a freeway in that time interval. VHT is represented in the units of vehicle-hours and can be calculated over a specific interval by the following equation.

$$\text{VHT} = \sum_i \frac{L \cdot q_i}{v_i} \tag{5}$$

Where, *v*, is the speed from the VDS in question.

Delay is represented in units of vehicle·hours and is calculated against a threshold speed.

$$\text{VHD} = \sum_i q_i \left( \frac{L}{v_i} - \frac{L}{b} \right) \tag{6}$$

Where, *v*, is the speed from the VDS in question and, *b*, is the threshold speed. The 65 mph threshold speeds are considered in this research for the freeway mainline. The two main chances for errors to arise in this method are:

1. usage of the g-factor approximation to estimate speeds
2. usage of a point measurement to approximate the measurement across an expanse of road

Single loops, as predominantly deployed, do not measure speeds, and one may expect that direct measurement of travel-times that are possible with third-party data can be advantageous.

**Hybrid Method**

The hybrid method is illustrated using **Figure 4** which shows input data, intermediate analysis method, and output with the estimated performance measures. The input data include flow and speed from the VDS and travel time (TT) data provided by third-party vendors. For the analysis of this research, one vendor is considered (Vendor A) who provides data in a separate spatial reference system that does not match that used by public agencies to locate VDS. For the freeway mainline, **Figure 4** shows such a situation where the links with travel time data do not align with the VDS locations. In this paper, a link refers to a length of the freeway for which travel time data is available from a third-party vendor. Also, the travel time data is not based on the whole vehicle population, it is from a sample probe vehicle group (x% of the whole population). Flow is conflated or projected to the desired cells along the freeway. Based on the conflated flow, travel time data from third-party vendors are also conflated. Once both flow and travel time data are available on the desired cells, the final performance measures are calculated by aggregating data from all cells in the freeway. In this research, the cell-based calculation is conducted for the freeway mainlines.

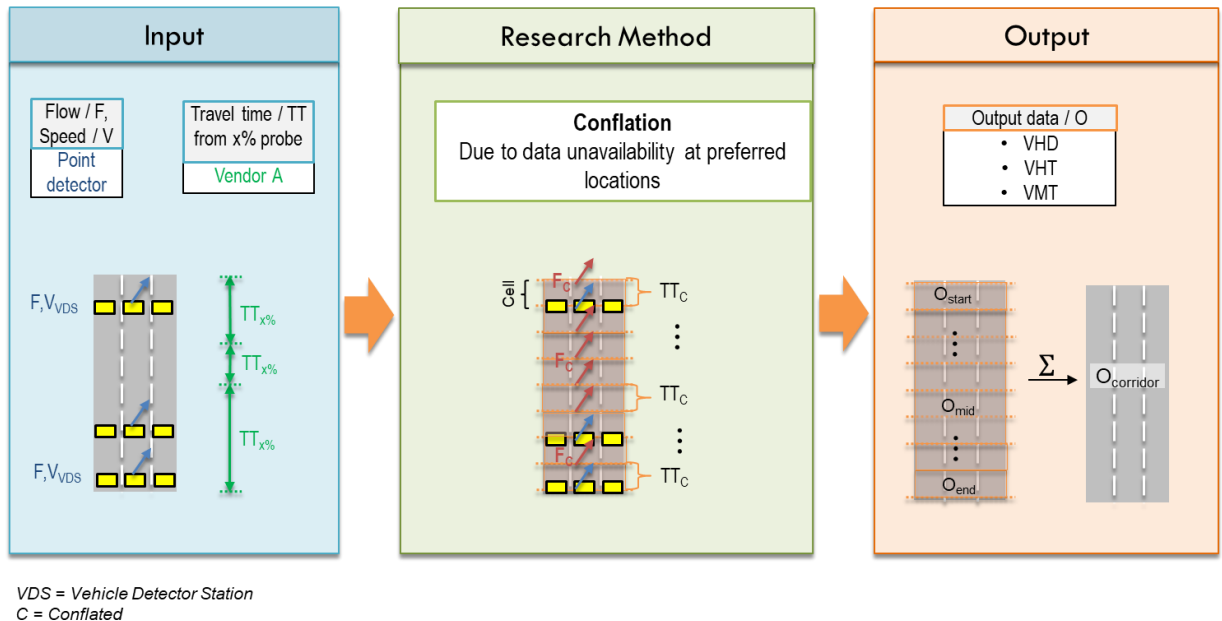

VDS = Vehicle Detector Station
C = Conflated

**Figure 4 Hybrid method to estimate performance measures**

*Hybrid Data Conflation*

In this subsection, the data conflation algorithm for the hybrid method is discussed. The main purpose of this step is to make both flow and travel time data from multiple data sources available on a single spatial reference system, in this case, cells along the freeway.

**Desired Cells of Analysis: Figure 5** shows a schematic of evenly-sized cells along the freeway mainline. In general, travel time data on links do not line up with VDS data on segments, which do not line up with the cells along the desired domain of analysis. The cells are the blue bounding boxes, and they cover freeway mainlines. The motivation of cell-based analysis is to narrow down the locations of the bottlenecks and compute delay properly using conflated flow in each cell. With large cells, variations in flow can not be properly captured and it can lead to erroneous delay calculation. In this analysis, the cell length is considered to be 0.25 mile. Therefore flow data from the VDS are projected every 0.25 miles. If any VDS is within 200 ft from a cell boundary, the cell location is not considered and the raw VDS data is used as-is. At the end of the conflation process, flow data is available at each VDS (measured flow) and cell boundary (conflated flow).

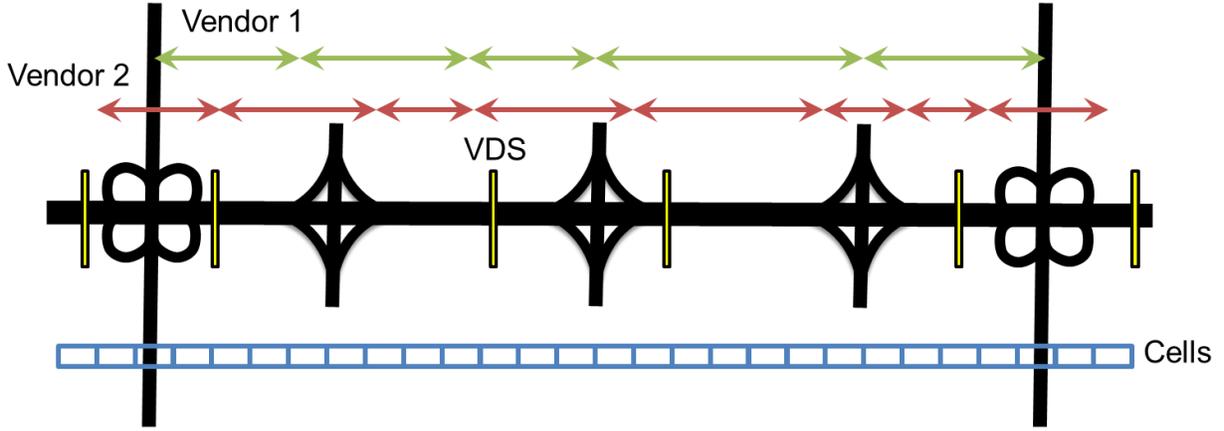

**Figure 5 Freeway with cells**

**Flow Conflation:** The flow conflation method is developed based on the GASM (Generalized Adaptive Smoothing Method) (5, 7). The purpose of the method is to spatio-temporally reconstruct traffic data at specific locations using data captured by point detectors or VDS. At certain cell points (where *x* and *t* are the position and time) on a spatio-temporal domain, the flow data *q* can be calculated using GASM. In this method, $f_{(x,t)}$ is a normalization factor, and *k* is the kernel value. According to GASM, the conflated flow at cell point *(x, t)* is obtained from all VDS captured flow values in the upstream and downstream regions. In GASM, localized smoothing is performed, meaning flow at a certain cell point *(x, t)* is affected strongly by the closer VDS, and weakly by the distant VDS. The widths of spatial and temporal smoothing are $\delta$ and $\mu$, respectively.

GASM is developed to overcome the challenge of isotropic smoothing of traffic data (i.e., non-skewing smoothing). The equations of the GASM method are provided here with the non-skewing smoothing. Here *i* and *j* refer to the time interval and VDS number, respectively, for the study period and analysis area. At any cell point located at *x* position, $q_{c(x,t)}$ is the flow at time interval *t*. The variable $q_{vds(i,j)}$ is the flow of the *j*-th VDS at the *i*-th time interval.

$$q_{c(x,t)} = \frac{1}{f_{(x,t)}} \sum_{i=1}^{T} \sum_{j=1}^{N} k_{(x-x_j, \ t-t_i)} \cdot q_{vds(i,j)} \tag{7}$$

$$f_{(x,t)} = \sum_{i=1}^{T} \sum_{j=1}^{N} k_{(x-x_j, \ t-t_i)} \tag{8}$$

$$k_{(x-x_j, \ t-t_i)} = \exp\left[-\left(\frac{|x-x_j|}{\delta} + \frac{|t-t_i|}{\mu}\right)\right] \tag{9}$$

Here *N* is the total number of VDS, and *T* is the last time interval. GASM includes the idea of skewed smoothing of traffic data. In the free-flow *(ff)* direction, the smoothing is performed with the freeflow propagation speed *(v_{ff})*. In the congested *(cong)* direction, the smoothing is performed with the backward propagation speed *(v_{cong})*. In this analysis, the available VDS are confined to the immediate upstream and downstream VDS, and thus the smoothing method is named as 'Confined Generalized Adaptive Smoothing Method (C-GASM)'. **Figure 6** shows the situation where intermediate cells between two VDS are conflated with the data from the immediate upstream and downstream VDS.

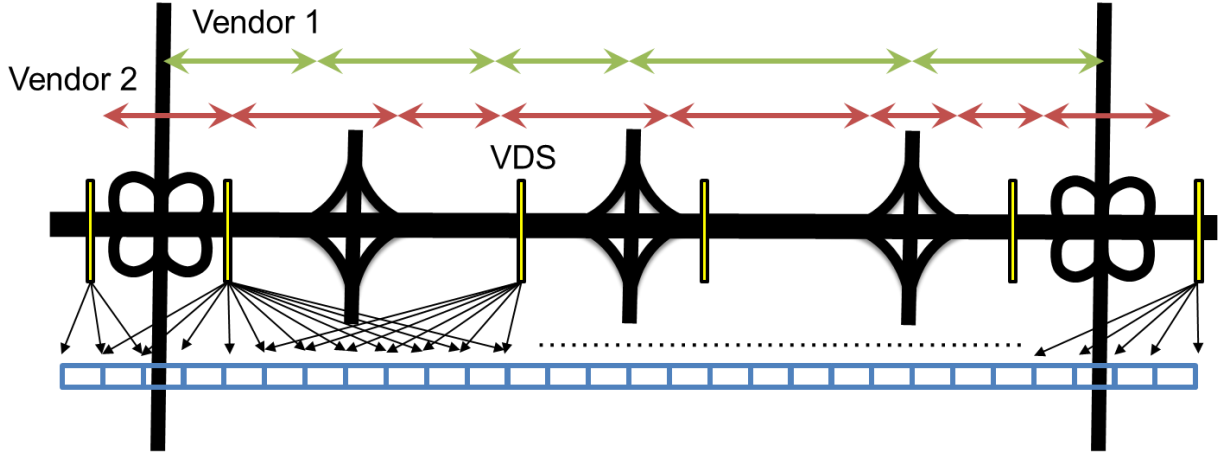

**Figure 6 C-GASM for conflating with surrounding VDS**

Based on the confinement rule, the values for conflated flow in congested and freeflow conditions at cell point *(x, t)* are estimated with **Equations 10** and **12**, respectively. Here *u* is the upstream VDS while *d* is the downstream one.

$$q_{c,ff(x,t)} = \frac{1}{f_{ff(x,t)}} \sum_{i=1}^{T} (k_{\left(x-x_u, t-t_i-\frac{x-x_u}{v_{ff}}\right)} \cdot q_{vds(i,u)} + k_{\left(x-x_d, t-t_i-\frac{x-x_d}{v_{ff}}\right)} \cdot q_{vds(i,d)}) \tag{10}$$

$$f_{ff(x,t)} = \sum_{i=1}^{T} (k_{\left(x-x_u, t-t_i-\frac{x-x_u}{v_{ff}}\right)} + k_{\left(x-x_d, t-t_i-\frac{x-x_d}{v_{ff}}\right)}) \tag{11}$$

$$q_{c,cong(x,t)} = \frac{1}{f_{cong(x,t)}} \sum_{i=1}^{T} (k_{\left(x-x_u, t-t_i-\frac{x-x_u}{v_{cong}}\right)} \cdot q_{vds(i,u)} + k_{\left(x-x_d, t-t_i-\frac{x-x_d}{v_{cong}}\right)} \cdot q_{vds(i,d)}) \tag{12}$$

$$f_{cong(x,t)} = \sum_{i=1}^{T} (k_{\left(x-x_u, t-t_i-\frac{x-x_u}{v_{cong}}\right)} + k_{\left(x-x_d, t-t_i-\frac{x-x_d}{v_{cong}}\right)}) \tag{13}$$

To calculate a single smoothed flow value for *(x, t)*, a weighted filter is used. With (, the final flow value at the cell point *(x,t)*, which is $q_{f(x,t)}$, is estimated.

$$q_{f(x,t)} = z_{(x,t)} \cdot q_{c,cong(x,t)} + (1 - z_{(x,t)}) \cdot q_{c,ff(x,t)} \tag{14}$$

The weight $z_{(x,t)}$ is calculated with an s-shape function, which depends on crossover speed ($v_{cr}$) and transition width ($\Delta v$) from congestion to free flow.

$$z_{(x,t)} = \frac{1}{2} \cdot [1 + \tanh(\frac{v_{cr} - \min(v_{c,ff(x,t)}, v_{c,cong(x,t)})}{\Delta v})] \tag{15}$$

At the cell point *(x,t)*, values of $v_{c,ff(x,t)}$ and $v_{c,cong(x,t)}$ are calculated using equations similar to **Equations 10 and 12** with the VDS-captured speeds. C-GASM based flow conflation method depends on parameters such as $\delta, \mu, v_{ff}$, $v_{cong}$, $v_{cr}$, and $\Delta v$. The typical values of these parameters are discussed in (*5, 7*). In this analysis, these parameters are selected based on multiple trials where the chosen set of acceptable values gives the highest accuracy. **Figure 7** shows an overview of the flow conflation method using C-GASM.

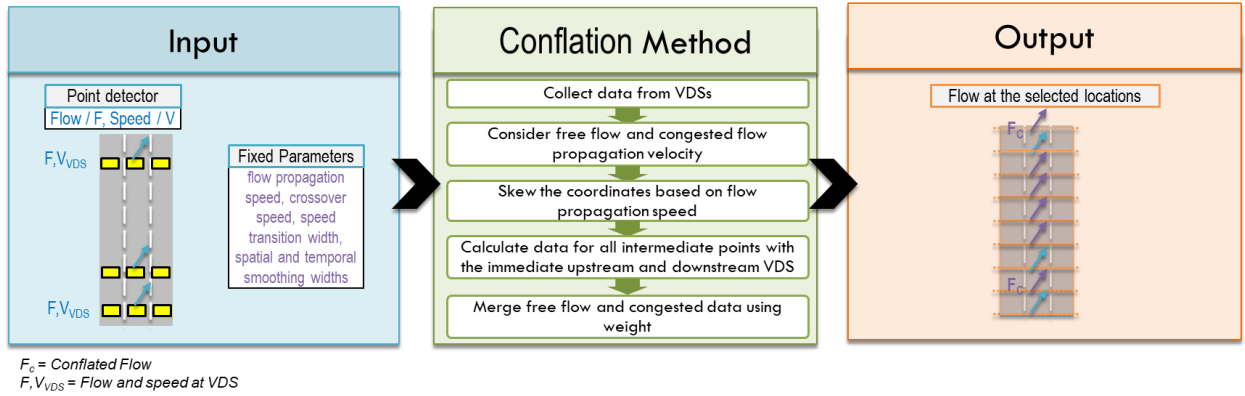

**Figure 7 C-GASM method**

**Third-party Data Conflation:** Once the flow values from the point sensors are conflated to the desired cell locations, the next step is to conflate the third-party data on the same cells. Travel time data provided by a third-party is conflated to the desired cells, as shown in **Figure 8**, where the link is divided by the overlapping cells. The assumption of travel time conflation is that travel time data can be distributed along the links to the cells based on the vehicle number distribution in the cells. The higher number of vehicles in a cell will result in higher travel time, and vice versa. Assume that, vendor A provided travel time data for a link is $TT_i$ at time interval *i*. In that link, the total number of cells is *G* which divide the link into *(G+1)* parts. For a certain cell located at *x* on that link, the associated travel time data ($tt_{x,i}$) at time interval *i* from vendor A, is estimated with this equation.

$$tt_{x,i} = TT_i \cdot \frac{c_{x,i}}{\sum_{n=1}^{G+1} c_{n,i}} \tag{16}$$

The number of vehicles in a cell ($c_{x,i}$) located at *x* for the time interval *i* can be estimated by the conflated density on the cell, and associated length of the link from that cell to the next cell. The conflated density is estimated with the C-GASM method.

Travel time for the cells, which cover edges of multiple links, are calculated by aggregating travel time for those link edges at time interval *i*. Travel time data from multiple vendors can be estimated with a weighted sum approach. The weight ($\emptyset$) can be assigned based on the confidence on the third-party vendor provided data. The confidence can be related to the travel time data characteristics (penetration level of probe vehicles, real-time data availability) of the vendor provide data.

$$tt_{x,i} = \emptyset_A \cdot tt_{A,x,i} + \emptyset_B \cdot tt_{B,x,i}, \emptyset \in [0,1], \emptyset_A + \emptyset_B = 1 \tag{17}$$

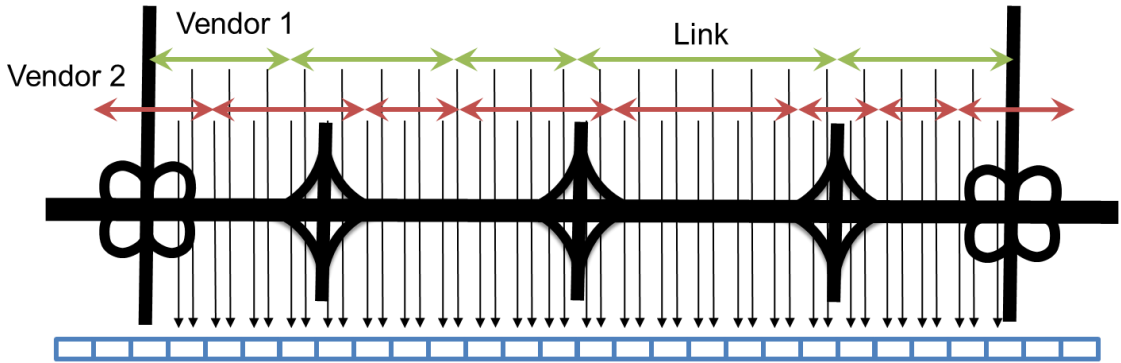

**Figure 8 Travel time conflation**

If multiple vendors have the same weights, ∅ will have equal values, and the sum of all weights will be equal to 1. In this research evaluation, only one vendor is considered.

*Hybrid Performance Measures Estimation*
Once multiple data from different sources are conflated on the same network along the desired cells, speed is estimated from the third-party travel time. Using **Equations 4-6**, VMT, VHT, and VHD are calculated for each cell. Final values for the freeway are calculated by summing up the values for the individual cell, as shown in **Figure 4**.

**Experimental setup**

To evaluate the performance of the traditional and hybrid methods, an experimental setup is developed and used with a simulated model of the I-210 corridor. **Figure 9** shows the calibrated model of the I-210 corridor that is used in this research. The simulation model is developed for the Connected Corridors program, which has different roadways (freeways, ramps, and arterials) calibrated for both weekends and weekdays (*18*).

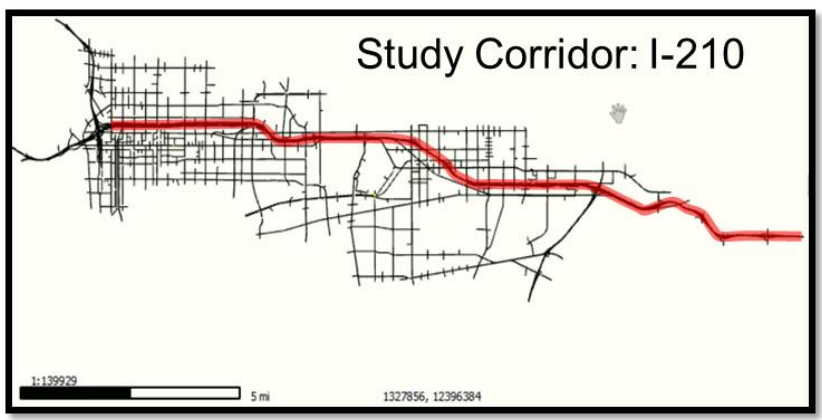

**Figure 9 I-210 simulation model**

The red highlighted freeway used in this analysis is the westbound portion of the I-210 freeway. The VDS in the simulated model are laid out following the VDS placement on the real-world freeway. To synthesize the third-party vendor travel time data, raw location data is collected from a sample of the total vehicles (e.g., 5% of the simulated vehicles). The probe vehicle data is only considered when the vehicle data is available for the initial and last 10% part of the link. This is done to ensure that the vehicle has actually crossed the link. From the initial and final location and timestamp data of the associated probe vehicle, travel time is calculated for that vehicle. Later, the data is aggregated for every minute time interval, and the final dataset has the travel time data aggregated for every minute. **Figure 10** shows the representation of how trajectories are used to calculate travel times. The probe vehicle data from P1, P2, and P3 are captured only when these vehicles crossed the section.

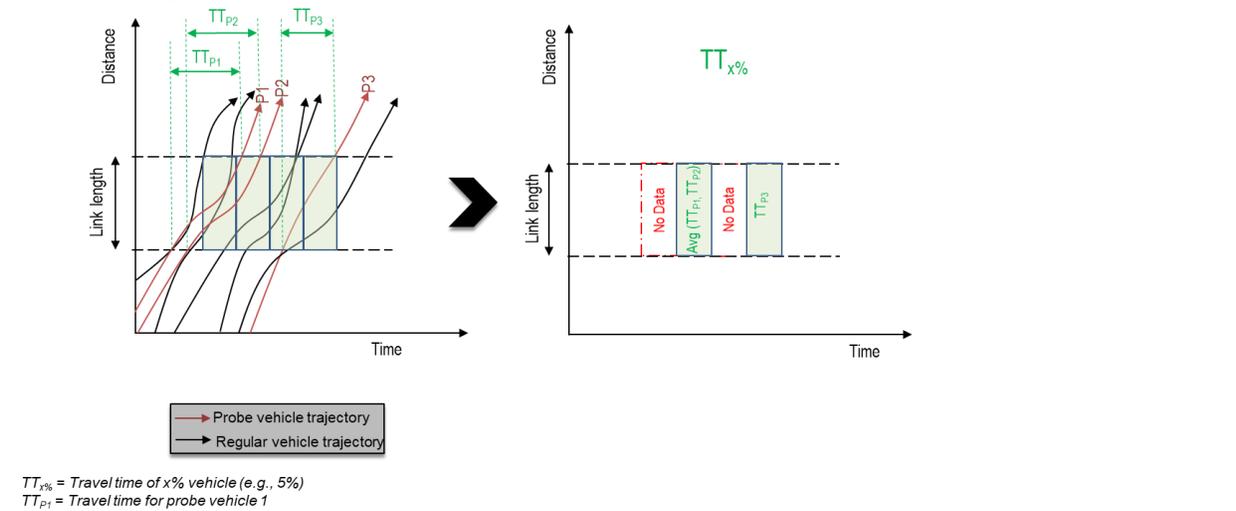

$TT_{x\%}$ = Travel time of x% vehicle (e.g., 5%)
$TT_{P1}$ = Travel time for probe vehicle 1

**Figure 10 Probe vehicle travel time data generation**

The 16-mile length of the westbound I-210 mainline lanes is used for the evaluation of the traditional and hybrid methods. Along the selected freeway, 33 VDS are available. The final number of cells along the freeway is 55. VHT and VHD are calculated with respect to the 65 mph speed threshold. VMT, VHT, and VHD are calculated for the following scenarios:

1. Before morning peak (6 am - 7 am)
2. Morning peak (7 am - 8 am)
3. Noon time (1 pm - 2 pm)
4. Afternoon peak (5 pm - 6 pm)
5. Night off-peak (8 pm - 9 pm)

The findings presented later are based on the average of two replications (i.e., different simulation runs using different random seed) for each of the scenarios.

**RESEARCH CONSIDERATIONS**

There are several assumptions considered in the research. For the input data, a one-minute time window is considered. This means data from VDS and a third-party are sampled at intervals of one minute. This time interval can be re-sampled to any other preferred time interval. In the analysis, travel time from a third-party is used for freeway mainlines, whereas, in reality, the data can include vehicles on other lanes too (e.g., HOV lanes).

**ANALYSIS AND FINDINGS**

In this section, findings from the point-detector based speed estimation, data conflation, and hybrid data fusion are discussed.

**Traditional Speed Calculation**

In order to calculate speed from single loop detectors available in the study area, the g-factor based speed estimation method is used. These g-factors, which are the average length of vehicles crossing a detector, influence the final calculated speed. A set of g-factors are used for each lane to identify which factor gives an acceptable range of error for almost all single loops. For a one-hour time interval, where loop data are aggregated for each minute, **Figure 11** shows the error of calculated speed (i.e., speed using g-factor) and actual speed (i.e., speed from the simulation) for a specific g-factor (i.e., 22 ft.) for all loops in lane 1 of the freeway mainlines. The error is considered to be acceptable if: (i) the 50th percentile value of the error range is close to 0, and (ii) the sample size is near-equal for the overestimated and underestimated values.

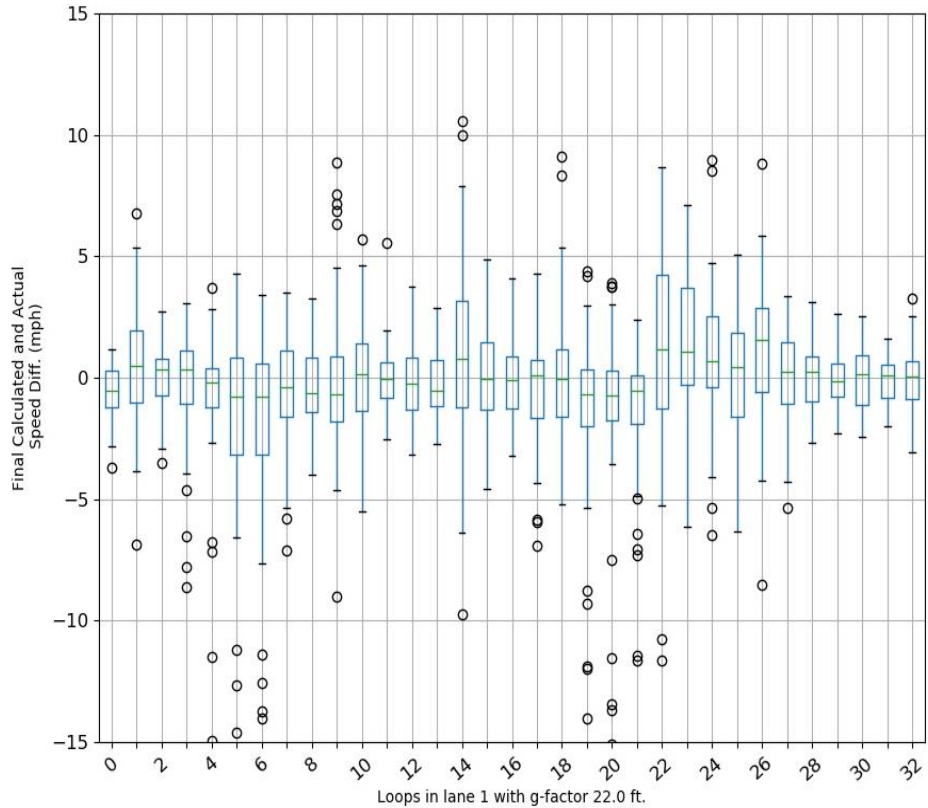

**Figure 11 Calculated speed error for lane 1 in freeway mainline**

For the freeway mainline, the final g-factor values for six lanes are found to be 22, 22, 26, 25, 24, and 23 ft., from the leftmost to the rightmost lane. **Table 1** shows the findings for the absolute difference between g-factor estimated speed and the simulated speed for the morning congested scenario. Here data are aggregated for all single loops in the freeway mainline. For the congested scenario, the mean speed difference is 2.32 mph with a standard deviation of 2.89 mph, which means that in simulation, the g-factor method can generate speed values that are very close to the simulated speeds.

**TABLE 1 Traditional Speed Estimation Result**

| Simulation Scenario | Absolute Difference of Final Calculated and Actual Speed (mph) | | | | | |
|---|---|---|---|---|---|---|
| | Sample | Mean | Std. deviation | 25th Percentile | 50th Percentile | 75th Percentile |
| Morning congestion | 8496 | 2.32 | 2.89 | 0.68 | 1.49 | 2.85 |

**Freeway Mainline Performance Measure**

This sub-section discusses the findings of both traditional and hybrid methods to calculate performance measures.

*Conflation for the Hybrid method*

Data conflation projects data (from different sources) onto the same spatial reference system. **Figure 12** shows a sample representation of the flow data availability from VDS along the westbound portion of I-210. If there is no VDS, no flow data is available and these regions are shown with white areas in **Figure 12**. The white areas represent the cells. The VDS captured flow data are conflated to the desired cells with the length of 0.25 miles. After applying the flow conflation using GASM and C-GASM methods, conflated flow at these cells are available. In the simulation model, additional detectors are placed in the cell locations only to measure the accuracy of the flow conflation methods.

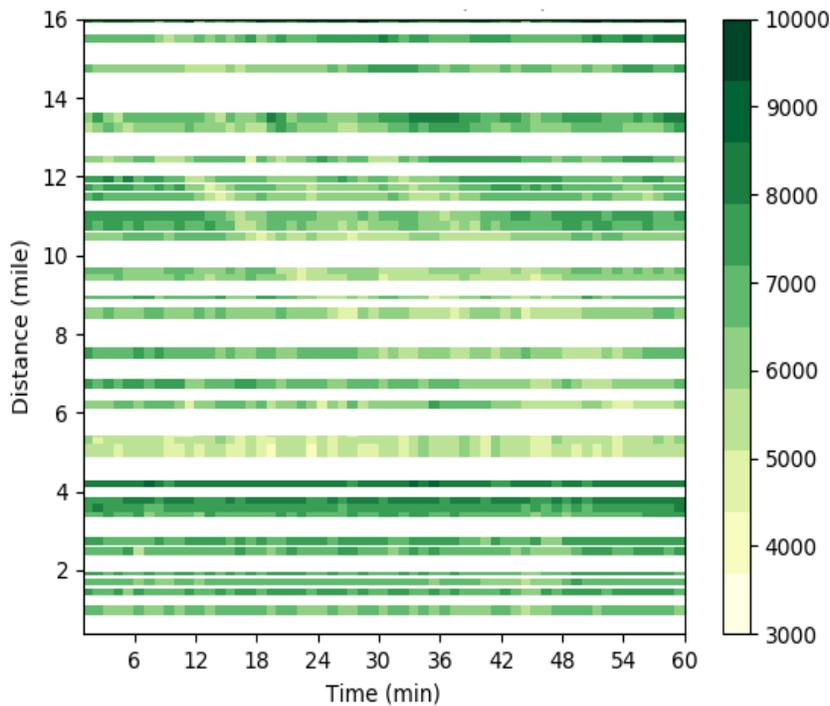

**Figure 12 Sample available data from VDS along I-210 westbound**

After multiple trials, the values of $\delta, \mu, v_{ff}, v_{cong}, v_{cr},$ and $\Delta v$ are found to be 0.8 km, 1 min, 100 kmph, 10 kmph, 90 kmph and 20 kmph, respectively, for the study area. **TABLE 2** shows the mean absolute error and mean absolute percentage error for every minute at all cell locations for the morning peak period. In the analysis, the error values are the average error of two simulated replications. As C-GASM exhibited superior performance during this heavily congested peak period, it is used in the final application of performance measure estimation.

**TABLE 2 Error of Flow Conflation at Morning Peak**

| Conflation Method | Mean Absolute Error (veh/hr) | Mean Absolute Percentage Error (%) |
|---|---|---|
| C-GASM | 482 | 8 |
| GASM | 515 | 8.2 |

*Comparison of Traditional and Hybrid Methods*

In the hybrid method, once both flow and travel time data are conflated, speed is estimated and the performance measures (VMT, VHT, and VHD) are calculated at the desired cell locations. Finally, all values along the whole freeway are summed up to get the final VMT, VHT, and VHD for the whole freeway.

**Table 3** shows the VMT, VHT, and VHD values for all scenarios, and methods. Simulated ground truth data is calculated from the model. The simulation provides space-mean speed and vehicle count data for the simulated sections, which are used to get the base VMT, VHT, and VHD. For each scenario, VMT calculated with the hybrid method is closer to the simulation ground truth, than that calculated with the traditional method. A small VHD is observed during the night off-peak scenario as an artifact of the cell-based travel time conflation, however, this can be considered as negligible.

**TABLE 3 Performance Measures for Freeway Mainline**

| Scenario | Calculation Method | VMT (veh-mile) | VHT (veh-hour) | VHD (veh-hour) |
|---|---|---|---|---|
| Before Morning Peak | SGT* | 94598.56 | 2924.55 | 1374.44 |
| | Traditional | 92624.45 | 2788.03 | 1270.74 |
| | Hybrid | 93316.80 | 2918.89 | 1456.94 |
| Morning Peak | SGT | 80237.48 | 4689.71 | 3354.89 |
| | Traditional | 77405.69 | 4338.11 | 3051.96 |
| | Hybrid | 78128.70 | 4630.18 | 3366.20 |
| Noon Time | SGT | 93634.13 | 2598.23 | 1064.56 |
| | Traditional | 92035.59 | 2641.62 | 1119.39 |
| | Hybrid | 92675.72 | 2507.76 | 1054.24 |
| Afternoon Peak | SGfT | 91021.54 | 3235.14 | 1743.29 |
| | Traditional | 89291.75 | 2975.35 | 1514.38 |
| | Hybrid | 90060.30 | 3098.24 | 1696.65 |
| Night Off-peak | SGT | 56544.19 | 822.31 | 0.04 |
| | Traditional | 55889.45 | 837.67 | 0.00 |
| | Hybrid | 56212.10 | 741.84 | 1.14 |

*\* SGT = simulated ground truth*

**Figure 13** shows the benefit of including third-party data for VMT, VHT, and VHD calculations. For the morning and afternoon peaks, the hybrid method yields an improvement of 9% and 10.4%, respectively, for VHD compared to the traditional method. Due to C-GASM based conflation, VMT also improves when using the hybrid method. Both traditional and hybrid methods underestimate all the performance measures in both scenarios.

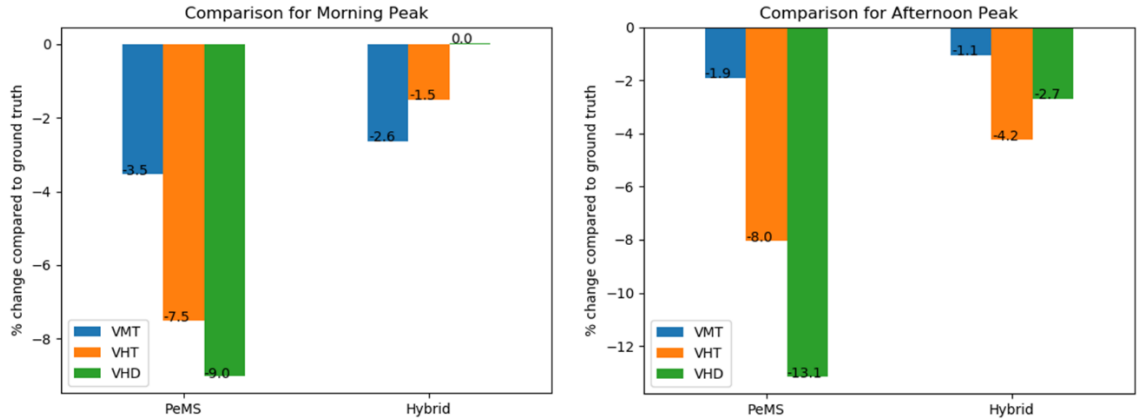

**Figure 13 Peak period comparisons for freeway mainline**

For the before-morning and noon scenarios, the hybrid method yields an improvement of 1.5% and 5.0%, respectively, for VHD compared to the traditional method as shown in **Figure 14**. Due to C-GASM based conflation, VMT also improves when using the hybrid method. For the noon scenario, traditional overestimates both VHT and VHD, whereas the hybrid method underestimates them.

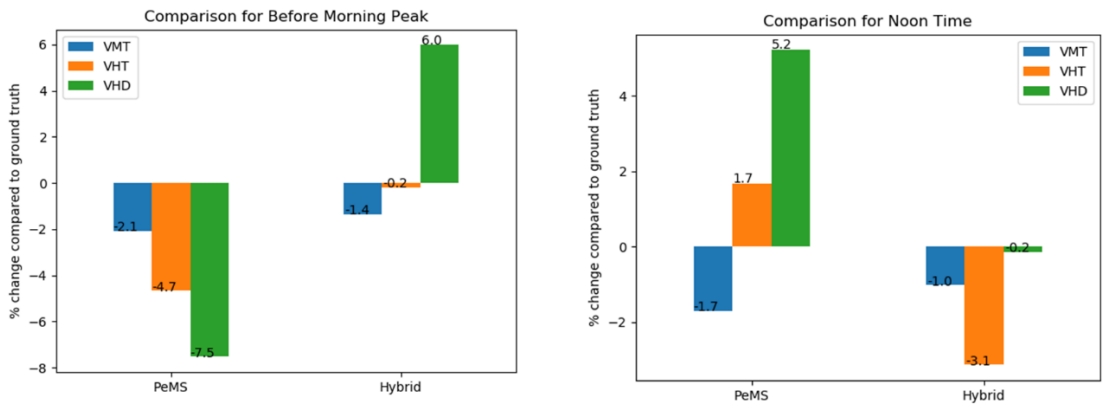

**Figure 14 Other period comparisons for freeway mainline**

**CONCLUSIONS**

The objectives of this research are to develop and evaluate a framework to calculate performance measures for freeway mainlines using a mix of data from multiple data sources, which include point detector and third-party vendors. The purpose is to enable public transportation agencies to check if third-party provided data can augment already existing data from the point detectors. A literature review is conducted to study the current practices of data conflation and data fusion. The experimental design includes the use of the I-210 simulated freeway. Research is conducted for freeway mainlines using the simulated model. Both g-factor based point speed estimation and limited probe data availability are considered to mimic real-world conditions.

Based on the experiment for morning congestion, the g-factor based speed calculation can generate a good estimation of point speed in simulation. However using the point speed to calculate freeway-wide VMT, VHT and VHD can produce erroneous results. For freeway mainlines, fusing data from third-party vendors helps to get a better estimation of the performance measures in almost all scenarios. For the off-peak period, when there is no noticeable fluctuation in demand, only point detector based estimation is enough to estimate the performance measures. This happens as the point-speed at off-peak periods is very close to space mean speed for the link.

In this study, the HOV lanes are not considered while estimating the freeway performance measures. Also, the effects of the historical travel time data (in case of no probe vehicle presence on the link) for the third-party vendors are not captured. Future studies should be conducted to address these limitations.

## ACKNOWLEDGMENTS

This work would not have been possible without the generous support from the California Department of Transportation. We acknowledge the strong support of Brian Simi and Jeff Forester for ongoing work on hybrid data and performance measurement.

## AUTHOR CONTRIBUTIONS

The authors confirm contribution to the paper as follows: study conception and design: Sakib Mahmud Khan. Anthony Patire; data collection: Sakib Mahmud Khan; analysis and interpretation of results: Sakib Mahmud Khan. Anthony Patire; draft manuscript preparation: Sakib Mahmud Khan. Anthony Patire. All authors reviewed the results and approved the final version of the manuscript.

**REFERENCES**

1.  Caltrans. Quarterly Reports | Caltrans. https://dot.ca.gov/programs/traffic-operations/mpr/quarterly. Accessed Jun. 7, 2020.
2.  Caltrans. PeMS Data Source | Caltrans. https://dot.ca.gov/programs/traffic-operations/mpr/pems-source. Accessed Jun. 6, 2020.
3.  Chen, X. Z. C. V. D. G. E. and M. *Practices on Acquiring Proprietary Data for Transportation Applications*. 2019.
4.  Bayen, A. M., P. E. Sharafsaleh, and A. D. Patire. *Hybrid Traffic Data Collection Roadmap: Objectives and Methods*. 2013.
5.  Treiber, M., A. Kesting, and R. E. Wilson. Reconstructing the Traffic State by Fusion of Heterogeneous Data. *Computer-Aided Civil and Infrastructure Engineering*, 2011. https://doi.org/10.1111/j.1467-8667.2010.00698.x.
6.  Van Lint, J. W. C., and S. P. Hoogendoorn. A Robust and Efficient Method for Fusing Heterogeneous Data from Traffic Sensors on Freeways. *Computer-Aided Civil and Infrastructure Engineering*, 2010. https://doi.org/10.1111/j.1467-8667.2009.00617.x.
7.  Treiber, M., and D. Helbing. Reconstructing the Spatio-Temporal Traffic Dynamics from Stationary Detector Data. *Cooper@tive Tr@nsport@tion Dyn@mics*, 2002.
8.  Ottaviano, F., F. Cui, and A. H. F. Chow. Modeling and Data Fusion of Dynamic Highway Traffic. *Transportation Research Record*, Vol. 2644, 2017, pp. 92–99. https://doi.org/10.3141/2644-11.
9.  Li, M., X. (Michael) Chen, and W. Ni. An Extended Generalized Filter Algorithm for Urban Expressway Traffic Time Estimation Based on Heterogeneous Data. *Journal of Intelligent Transportation Systems: Technology, Planning, and Operations*, 2016. https://doi.org/10.1080/15472450.2016.1153426.
10. Khaleghi, B., A. Khamis, F. O. Karray, and S. N. Razavi. Multisensor Data Fusion: A Review of the State-of-the-Art. *Information Fusion*, 2013. https://doi.org/10.1016/j.inffus.2011.08.001.
11. Liu, J., T. Li, P. Xie, S. Du, F. Teng, and X. Yang. Urban Big Data Fusion Based on Deep Learning: An Overview. *Information Fusion*, Vol. 53, 2020, pp. 123–133. https://doi.org/10.1016/j.inffus.2019.06.016.
12. Ambühl, L., and M. Menendez. Data Fusion Algorithm for Macroscopic Fundamental Diagram Estimation. *Transportation Research Part C: Emerging Technologies*, 2016. https://doi.org/10.1016/j.trc.2016.07.013.
13. Patire, A. D., M. Wright, B. Prodhomme, and A. M. Bayen. How Much GPS Data Do We Need? *Transportation Research Part C: Emerging Technologies*, 2015. https://doi.org/10.1016/j.trc.2015.02.011.
14. Wu, C., J. Thai, S. Yadlowsky, A. Pozdnoukhov, and A. Bayen. Cellpath: Fusion of Cellular and Traffic Sensor Data for Route Flow Estimation via Convex Optimization. *Transportation Research Part C: Emerging Technologies*, Vol. 59, 2015, pp. 111–128. https://doi.org/10.1016/j.trc.2015.05.004.
15. Wang, Y., Y. Zhang, Z. S. Qian, S. Wang, Y. Hu, and B. Yin. Multi-Source Traffic Data Reconstruction Using Joint Low-Rank and Fundamental Diagram Constraints. *IEEE Intelligent Transportation Systems Magazine*, Vol. 11, No. 3, 2019, pp. 221–234. https://doi.org/10.1109/MITS.2019.2919529.
16. Wright, M., and R. Horowitz. Fusing Loop and GPS Probe Measurements to Estimate Freeway Density. *IEEE Transactions on Intelligent Transportation Systems*, 2016. https://doi.org/10.1109/TITS.2016.2565438.
17. Zwet, E. van, C. Chen, Z. Jia, and J. Kwon. A Statistical Method for Estimating Speed from Single Loop Detectors. https://drive.google.com/file/d/0B5wZ4dLpgONnT1NxbG9SQ21rR1k/edit. Accessed Jun. 14, 2020.
18. Connected Corridors. I-210 Pilot Landing Page | Connected Corridors Program. https://connected-

corridors.berkeley.edu/i-210-pilot-landing-page. Accessed Jun. 17, 2020.